\documentclass[8pt,a4paper,twocolumn]{article}
\usepackage{amsmath,amsfonts,amssymb}
\usepackage{graphicx}
\usepackage{adjustbox}
\usepackage{subfigure}
\usepackage{float}
\usepackage{color}
\usepackage{url}
\usepackage[utf8]{inputenc}
\usepackage{multicol}
\usepackage{pifont}
\usepackage{xcolor}
\usepackage{textcomp}

\date{\today}
\usepackage[superscript,biblabel]{cite}

\usepackage{amsmath,amsfonts,amssymb}

\newcommand{\SILiquidProperties}{Supplementary Table~S1}%
\newcommand{\SIDihedralAngles}{Supplementary Figure~S1}%
\newcommand{\SIEndToEndHistogram}{Supplementary Figure~S2}%
\newcommand{\SIRDFClassicalvsQuantum}{Supplementary Figure~S3}%
\newcommand{\SIDiffusionSubs}{Supplementary Figure~S4}%
\newcommand{\SIRDFReactiveSites}{Supplementary Figure~S5}%
\newcommand{\SIRDFlinker}{Supplementary Figure~S6}%
\newcommand{\SIDiffusionFreeMonomer}{Supplementary Figure~S7}%
\newcommand{\SIDegreeDistribution}{Supplementary Figure~S8}%
\newcommand{\SICycloAllMethyl}{Supplementary Figure~S9}%
\newcommand{\SITriangles}{Supplementary Figure~S10}%
\newcommand{\SILargestComponent}{Supplementary Figure~S11}%
\newcommand{\SIHoles}{Supplementary Figure~S12}%
\newcommand{\SIAllTG}{Supplementary Figure~S14}%
\newcommand{\SIYoung}{Supplementary Figure~S15}%
\newcommand{\SIStrainRate}{Supplementary Figure~S16}%
\newcommand{\SIRes}{Supplementary Figure~S17}%
\date{}
\begin{document}

\twocolumn[  
\begin{@twocolumnfalse}

\title{Effect of different monomer precursors with identical functionality on the properties of the polymer network}
\author{Ariana Torres-Knoop$^1$,
 Verena Schamboeck$^2$, Nitish Govindarajan$^{2,3}$,\\ Pieter D. Iedema$^2$, and Ivan Kryven$^{4,5}$\footnote{i.kryven@uu.nl}\\
$^1$SURFsara, Science Park 140, 1098 XG  Amsterdam, the Netherlands\\
$^2$Van 't Hoff Institute for Molecular Sciences, University of Amsterdam, the Netherlands\\
$^3$Department of Physics, Technical University of Denmark, 2800 Kgs Lyngby, Denmark\\
$^4$Mathematical Institute, Utrecht University, Budapestlaan 6, 3508 TA Utrecht, the Netherlands\\
$^5$Centre for Complex Systems Studies, 3584 CE Utrecht, the Netherlands
}

\maketitle

\begin{abstract}
\noindent Thermo-mechanical properties of polymer networks depend on functionality of the monomer precursors -- an association that is frequently exploited in materials science. 
We use molecular simulations to generate spatial networks from chemically different monomers with identical functionality and show that such networks have several universal graph-theoretical properties as well as near universal Young's modulus. The vitrification temperature is shown to be universal only up to a certain density of the network, as measured by the bond conversion. The latter observation is explained by the fact that monomer's tendency to coil enhances formation of topological holes, which, when accumulated in the network, amount to a percolating cell complex restricting network's mobility. This higher-order percolation occurs late after gelation and is shown to coincide with the onset of brittleness, as indicated by a sudden increase in the glass transition temperature. This phenomenon may signify a new type of phase transition in polymer materials.\\
 \vspace{0.5cm}
\end{abstract}

\vspace{2cm}
\end{@twocolumnfalse}
]

\section*{Introduction}
Although it has been long hypothesised that structure of a covalently-bonded polymer network defines macroscopic physical properties of the material \cite{de1976relation}, little is known about what structural patterns are being formed in such networks and what parameters decide their fate. Polymerisation is driven by reaction kinetics and physical chemistry, and at the same time, the kinetics  is itself mediated by the evolution of the polymer network and its emergent geometry, which constitutes a complex paradigm for modellers. One of the classical results that dates back to the works of Flory and Stockmayer \cite{stockmayer1943theory,flory1969statistical,ziff1980kinetics} states that precursor functionality has a strong impact on the network structure and thus its properties. However, it remains unclear to what extent does one have a freedom to manipulate network structure when the precursor functionality is fixed.

\begin{figure}
\centering
\includegraphics[width=\columnwidth]{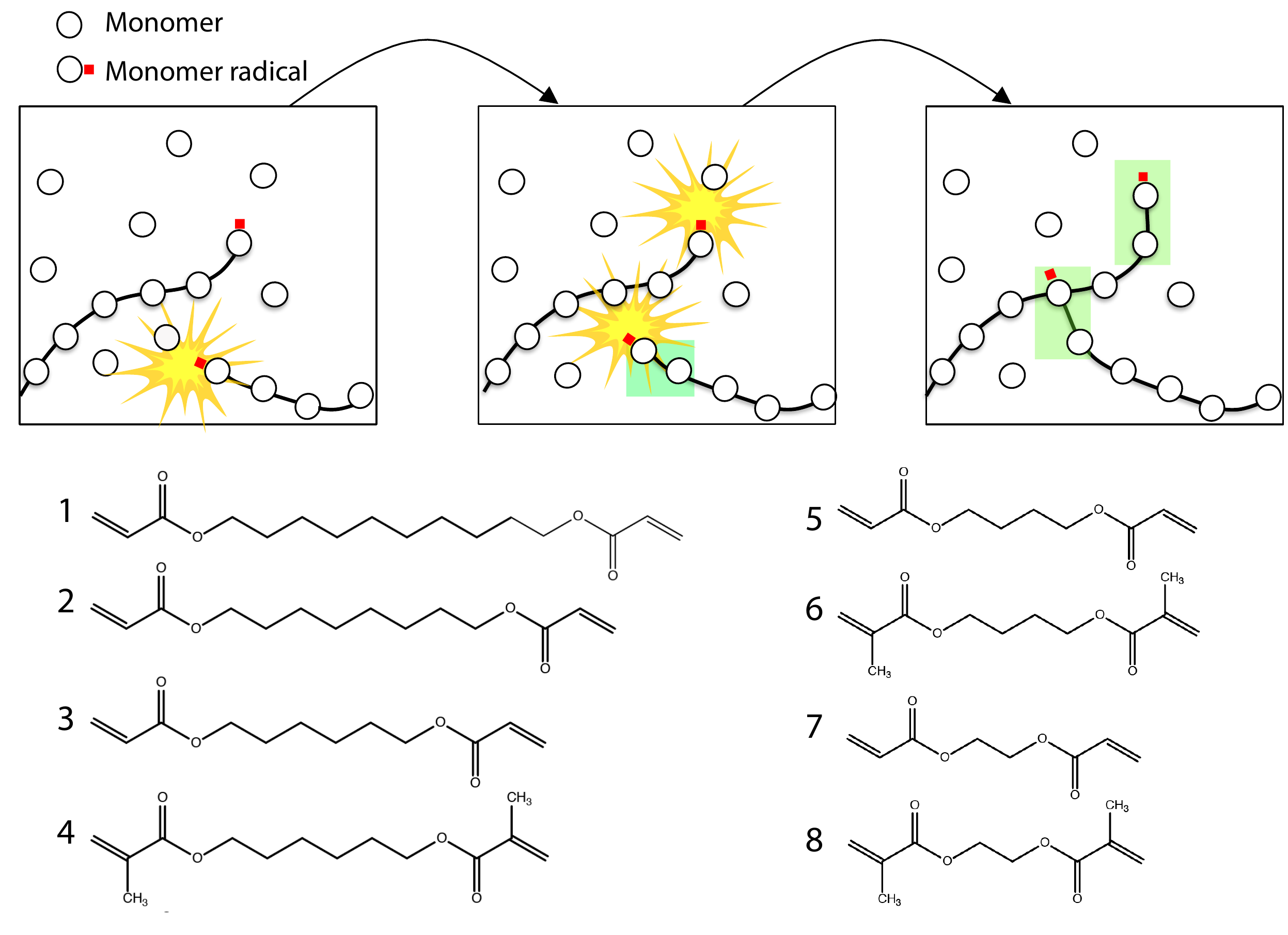}
\caption{{\bf Radical polymerisation of diacrylate monomers.} The network grows only at a few locations called \emph{monomer radicals} (labelled with {\color{red} \tiny$\blacksquare$}).
When reacting, a monomer radical and another node become connected. A node may obtain maximum 4 connections as it is identified with one of the following acrylate monomers: 
1) 1,10-decanediol diacrylate (DDDA),
2)1,8-octanediol diacrylate (ODDA),
3) 1,6-hexanediol diacrylate (HDDA),
4) 1,6-hexanediol dimethylacrylate (HDDMA),
5) 1,4-butyldiol diacrylate (BDDA),
6) 1,4-butyldiol dimethylacrylate (BDDMA),
7) 1,2-ethylenediol diacrylate (EDDA),
8)  1,2-ethylenediol dimethylacrylate (EDDMA).
}\label{fig:process}
\end{figure}

In this work, we use molecular simulations to perform a screening study of a range of monomers of identical functionality by varying the linker length and adding non-functional groups.
Namely, we study difunctional (meth)acrylates, see Figure \ref{fig:process}, that bear two double bonds allowing them to appear in the final network as nodes with 0, 1, 2, and 4 connections. Hence the only difference between the monomers lies in the length of the linker between the double bonds and/or the presence or absence of a bulky non-functional group close to the radical site (methyl substituent).  Standard theories that coarse-grain repeat units by ignoring their structural formulas postulate emergence of universal network properties for all the monomers  \cite{Kryven2016,Kryven2017,kryven2018,schamboeck2019}. Although such coarse-grained theories are irreplaceable when studying the rates of multiple competing reactions or tuning species concentrations, the purpose of this work is to put this hypothesis under scrutiny and investigate if subtle changes to nonreactive parts of monomers may nevertheless reflect on the structure of the network, and therefore, on the thermo-mechanical properties of the polymer.

Molecular dynamics (MD) may simulate geometrical and topological aspects of polymer networks allowing to study their effect on the thermo-mechanical properties \cite{rudyak2019thermoset,klahn2019effect,huang2019effects,demir2019versatile,jung2019free,Gissinger2020}.
Jung \emph{et al.}  \cite{jung2019free} used MD simulations to study the chain-growth polymerisation of the monofunctional methyl methacrylate (MMA) and reported volume shrinkage and glass transition temperatures.
Demir \emph{et al.}  \cite{demir2019versatile} and Huang \emph{et al.}  \cite{huang2019effects} employed MD to simulate copolymerisation of the vinyl ester/styrene polymer network. They reported structural polymer properties, such as the formation of cycles, the glass transition and Young's modulus.
Kl\"ahn \emph{et al.}  \cite{klahn2019effect} studied the effect of different monofunctional (meth)acrylate monomers on the glass transition temperature using molecular simulations.
Rudyak \emph{et al.}  \cite{rudyak2019thermoset} studied the effect of the initiator concentration on a chain-growth polymer network and reported gel conversions, gel fractions and cycle length distributions. 
In our previous study, we established a simulation protocol for polymerisation of 1,6-hexanediol diacrylate (HDDA), and observed the formation of `short cycles' during the polymerisation process: small cycles were created in primary and secondary cyclization reactions and were promoted by enhanced monomer flexibility \cite{TorresKnoop2018}.  On the one hand, the ability to form cycles is dictated by the local conformational dynamics of monomers, namely by the torsional strain and steric hindrance. On the other hand, these reactions influence the global properties of the network, such as the density and excluded volume by introducing topological defects that are elastically inactive. This has motivated us to enquire whether chemically different monomers with identical number of functional groups may lead to different gelation times and elastomeric properties of the network. Since the non-functional part of monomer precursors can be chemically manipulated,  the mechanism linking the monomer structure with the final polymer network is important for enabling rational material design.

The problem of optimising molecular weight, elastic modulus, or glass transition temperature arises in different contexts, for example, when designing self-healing materials  \cite{wang2020room}, but also in lithography, coatings for biomaterials, textiles and dental composites  \cite{bio1,lito,bio2,coating, coatings2,bio3,bio4}. Polymer resin recipes are designed to include multiple multifunctional monomer precursors to achieve a desired degree distribution \cite{schamboeck2019}. Acrylate polymers assembling into covalently bonded crosslinked networks with several vinyl groups per monomer  \cite{vanSteenberge2019} and can be tuned by adjusting different substituents in the $\alpha$-carbon, linkers between the vinyl groups, and by adding functional groups  \cite{gao2020complex}.  Additionally, manipulating chemical kinetics by adding a chain transfer or chain shuttling agents is used to reduce molecular weight \cite{vanSteenberge2019}, and similar techniques may be expected to reflect on the overall network structure, when used to form gel.

We start Results section, by reporting physical properties related to monomer diffusion, glass transition and elasticity. Then, we turn to the topological analysis of the network and its emergent geometry and match the structural changes in the network with the development of the physical properties. We show universality across the studied monomer types with respect to such properties as the degree distribution, cyclomatic number, gel point conversion, and Young's modulus.
We also report that such universality is broken in the dense networks, as can be seen by diverging glass transition temperatures associated with different monomer types. We show that this divergence may be mediated by small chordless cycles, also called topological holes, and report two qualitative transitions in the network structure. The first transition -- gelation, is marked by the appearance of an extensive cluster and is responsible for emergence of elastic behaviour. The second transition is marked by emergence of an extensive higher-order topological entity called a \emph{cell complex}, and is associated with a steep increase of glass transition temperature that may cause an onset of brittleness.

\section*{Results}
\subsection*{\label{ssec:poly_process}Polymerisation process}
During radical polymerisation, initially disconnected monomer units join together to form a growing network through three main reactions: initiation, propagation, and termination. 
As a matter of convention, we visualise these processes by representing monomers as labelled nodes of various functionality that receive connections according to rules shown in Figure \ref{fig:process}. However, the real radical polymerisation is a process that is also affected by the microscopic properties, \emph{e.g.}, monomer mobility and geometry, as well as by the macroscopic topological and thermo-mechanical changes in the whole system, \emph{e.g.}, viscosity, gelation, glass transition, volume and overall density. 
We study this process with dynamical systems that govern the motion of all atoms individually. The methods section explains the set-up of our curing simulations and the validation of the force field model.

\subsection*{Diffusion}
\begin{figure*}
\centering
\includegraphics[width=1.02\textwidth]{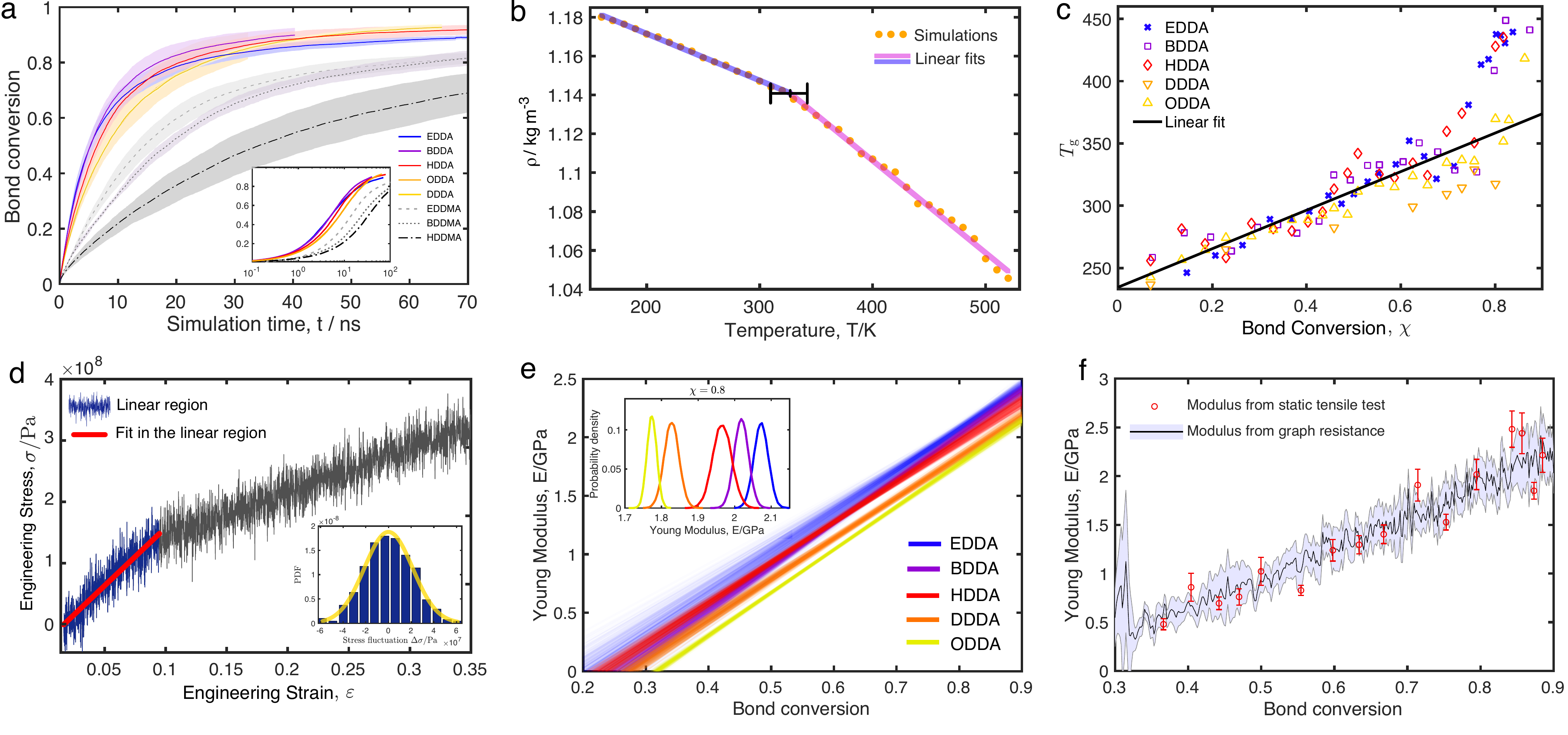}
\caption{{\bf Inferring free monomer diffusion and thermo-mechanical properties from molecular simulations}
({\bf a}) Bond conversion vs. simulation time. Confidence intervals indicate one standard deviation. Inset shows time in logarithmic scale.
({\bf b}) The procedure for calculating $T_g$ as the break point in a piecewise linear fit of density vs. temperature $\rho$, see Methods for details.
({\bf c}) $T_g$ vs. bond conversion features a universal collapse to a linear dependence. \SIAllTG~ reports confidence intervals for these data.
({\bf d})  The stress-strain curve with a thermal noise is processed to identify the region of initial linear growth and its slope, see Methods for details. 
The deviations from the fit, shown in the inset, are confirmed to be normally distributed and not autocorrelated.
({\bf e}) Young's modulus features a linear dependence on bond conversion resulting in fuzzy linear fits due to thermal fluctuations.
 The cross-sections of the probability densities of the fits at $\chi=0.8$  are shown in the inset.
  \SIYoung~reports confidence intervals. 
({\bf f}) Prediction of  Young's modulus from solely spectral graph properties of the network, is shown for the HDDA monomer.  See \SIRes\,for resistance-elastic moduli comparison for other monomers.  The elastic modulus and $T_g$ were averaged over four simulation runs.
}
\label{fig:Figure2}
\end{figure*}

In growing polymer networks, reactions rapidly become diffusion limited.
 Bond conversion $0\leq\chi(t)\leq1$ is the ratio between the formed covalent bonds at time $t$ and the maximal number of covalent bonds that the system may contain. Since each monomer can have maximum 4 neighbours, and each radical reduces the total number of bonds by 2, we have:
$$
	\chi(t)=\frac{b(t)}{2 N_\text{sys} - R_0},
$$
where $b(t)$ is the current number of bonds, $R_0=100$ the initial number of radicals, and $N_\text{sys}$ is the number of monomers in the system.  Figure~\ref{fig:Figure2}a quantifies the kinetic slowdown $\chi(t)$ for each monomer type, where we observe that: 
(1) for a given linker length, the presence of a methyl substituent significantly slows down bond formation, 
(2) for a given substituent, the polymerisation process slows down with increasing linker length, and
(3) the mobility of all systems slows down with curing. In \SIDiffusionSubs, we compare of the diffusion coefficients of free monomers  
as a function of conversion for systems with and without a methyl substituent. These results show that the decrease in the speed of polymerisation is  affected by the overall reduced translational mobility of bulkier dimethacrylates. One can also conclude from the radial distribution functions (RDFs) in the melt state, shown in \SIRDFReactiveSites,  that the addition of a methyl substituent reduces the frequency of reactions, as reactions involving the $\alpha$-carbon become less likely due to the steric effects. The analysis of the RDFs suggests that shielding a radical site by longer linkers is negligible, see \SIRDFlinker, which was also pointed out by Kurdikar and Peppas \cite{Kurdikar1994}. Therefore the diffusion of free monomers might play a dominant contribution to the overall kinetic slowdown, see \SIDiffusionFreeMonomer, which constitutes perhaps the strongest difference between the studied monomer types. 

The lower polymerisation rate of methacrylates as compared to acrylates for mono- and high-functional monomers is generally observed, both experimentally \cite{Buback2009,Wen1997} and computationally \cite{Yu2008}.
In previous studies on monofunctional acrylates, the decrease of the propagation rates with increasing linker length is often attributed to the increase of the activation energy. As we only observe effects of diffusion and steric hindrance, it is more accurate to compare the trend in Figure~\ref{fig:Figure2}a with reported trends on the pre-exponential factors of the rate coefficients. Mavroudakis \cite{Mavroudakis2015} reports the pre-exponential factors for monomers with varying linker lengths, which is in good agreement with our findings. Kurdikar and Peppas \cite{Kurdikar1994} investigated the effect of the linker length on the reaction kinetics of the polymerisation of different poly (ethylene glycol) diacrylates and found a decrease in the rate constant for longer linkers, attributing this effect to a decreased monomer diffusivity. 

After confirming that chemical reactions take place on different time scales for different monomers, in what follows we report all results as a function of bond conversion $\chi$ instead of time. In this way we focus on the outcome of the polymerisation and not merely on its absolute speed.

\subsection*{\label{ssec:glass}Thermomechanics}
 $T_g$ is determined by analysing the density as a function of  temperature  \cite{yang2016glass}.  As illustrated in  Figure~\ref{fig:Figure2}b and explained in Methods, we search for such a temperature, where this function fails to be smooth.  As shown in Figure~\ref{fig:Figure2}c,  $T_g$ increases with bond conversion for all monomers. 
 The increase is first linear and well-explained by a  \emph{universal}  relationship across all monomer species: $$T_\text{g} (\chi)= A\chi+T_\text{g,0},\; A=155 \text{K},$$  when $\chi<0.7$.
 The linear trend is abruptly broken at higher bond conversions. In the Topology section, we argue that a novel higher-order percolation  transition provides an explanation for such an abrupt change of $T_g$. The data presented in Figure~\ref{fig:Figure2}c suggest that shorter linkers might promote higher $T_g$ for $\chi>0.7$.
 
Both observations, the linear dependence and the deviation that follows it, are in line with findings in literature.
Bowman \emph{et al.} \cite{kannurpatti1998study} reported the $T_g$ for the homo- and copolymerisation of DEGDA and DEGDMA  with monofunctional monomers (n-octyl methacrylate and n-heptyl acrylate, respectively). These authors observed an increase in $T_g$ with increasing crosslinker density.
Bowman \emph{et al.} \cite{Bowman1990} also studied the effect of the distance between functional groups for poly (ethylene glycol) dimethacrylates from 1 to 4 ethylene glycol units and reported glass transition between 250 and 280 C. Kurdikar and Peppas \cite{Kurdikar1995}, reported the $T_g$ for the different poly (ethylene glycol) systems and reported that a longer linker reduces the $T_g$.
Kl\"an \emph{et al.} \cite{klahn2019effect} also observed lower $T_g$ for longer linkers in polymerized (meth)acrylates using molecular simulations.
  Jerolimov \emph{et al.} \cite{Jerolimov1994} studied the effect of divinyl crosslinking agents with different chain length (EGDMA and TEGDMA) on the $T_g$ of PMMA and also reported a similar observation. 
Recently, Kl\"an \emph{et al.} \cite{klahn2019effect} observed that  $T_g$ increases in the presence of the methyl group using molecular simulations.

\subsection*{\label{ssec:young}Elasticity}
Young's modulus is determined by estimating the slope of the stress-strain curve in the linear region, when a sample is pulled at a constant rate in the static tensile test. We perform such test under conditions of thermal fluctuations, see Figure~\ref{fig:Figure2}d, and as explained in Methods, statistical tests were applied to infer the linear region and its slope.

Figure \ref{fig:Figure2}e shows that for all monomer systems Young's moduli increase linearly as a function of bond conversion:
 $$E(\chi) = B(\chi-\chi_0) ,$$
 featuring a universal slope $B=3.6\text{GPa}$, with $\chi_0$ being the largest $\chi$, for which $E=0$.  Parameter $\chi_0$ turns out to be weakly dependant on linker length: Monomers with longer linker length feature smaller values of the moduli with an exception of HDDA-BDDA pair that deviates from the universal behaviour for $\chi<0.6$ showing a reverse order. The overall effects of the crosslinker length and bond conversion on elasticity are in agreement with experimental findings \cite{cook1990influence,davis1988properties}.
 
The evolution of the elastic modulus is universally explained for all monomers also by a spectral property of the networks called \emph{graph resistance}, see Methods for calculations. Figure~\ref{fig:Figure2}f compares the modulus obtained from the tensile test and the scaled graph resistance, for example of HDDA. As shown in \SIRes, presenting this comparison for other monomers, the graph resistance has to be scaled by a constant that absorbs the contribution from a single monomer coil and ranges between 5.0 and 8.8GPa.

\begin{figure*}
\includegraphics[width=\textwidth]{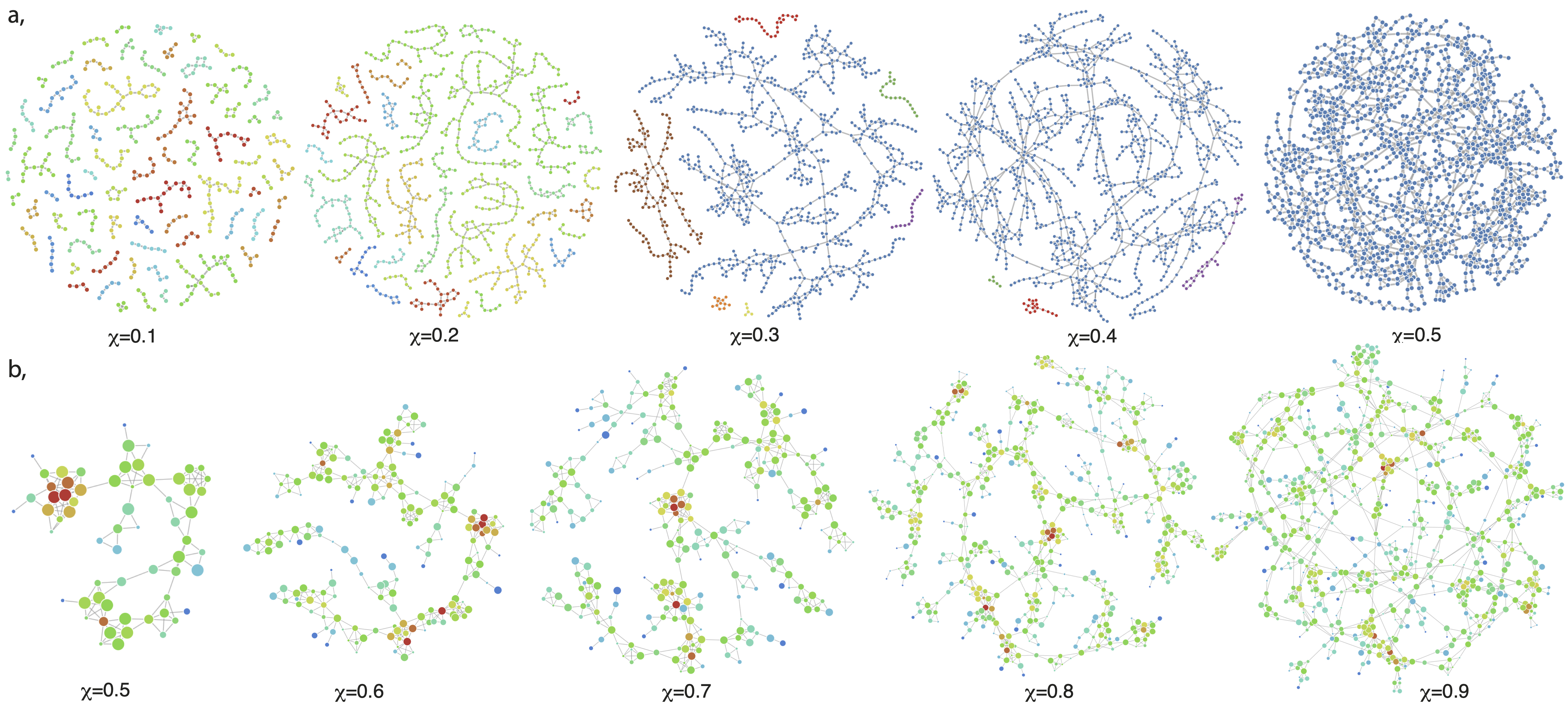}
\caption{
 {\bf Two-dimensional layouts of networks and cell-complexes developing in polymerising DDDA.}
({\bf a}) Network structures with highlighted components. The first percolation transition is thought to take place at $\chi=0.2$ and shortly afterwards the whole system predominantly consists of one connected component (isolated nodes not shown). The largest component continues to evolve as new bonds appear internally. ({\bf b}) Evolution of the largest connected component in the cell complex. Nodes represent holes in the former network: node size represents the size of the hole, with smallest nodes corresponding to size 3 and the largest to size 6. The second percolation transition is thought to take place at $\chi=0.7$, wherein the whole cell complex  becomes spanned by a component of an extensive size, which severely restricts mobility of the polymer network.  }
\label{fig:simplex}
\end{figure*}

\subsection*{\label{ssec:topology}Topology}
As a result of polymerisation, monomers join together with covalent bonds forming dense polymer networks, for example, as the series shown in Figures~\ref{fig:simplex}a for the DDDA system.
So far, we have discussed emergent physical properties of the polymer related to monomer diffusion, density, and elasticity. For all of these quantities, with an exception of the monomer diffusion coefficient, the number of bonds was a more important determining factor than the chemical type of the monomer. This observation suggests that the physical properties under investigation are mainly defined by the network structure, and only indirectly, by the monomer type. In other words: the network structure is a necessary intermediary between the monomer structure into the macroscopic physical properties. 

We quantify the structure of polymer networks in terms of several notions used in topology and (spectral) graph theory. Namely, we look at degree distribution, graph resistance, connected components, cycles, holes, and cell complexes. We analyse the dependence of these quantities on the monomer type, and argue about their effect on the emergent physical properties.

\begin{figure*}
\centering
\includegraphics[width=\textwidth]{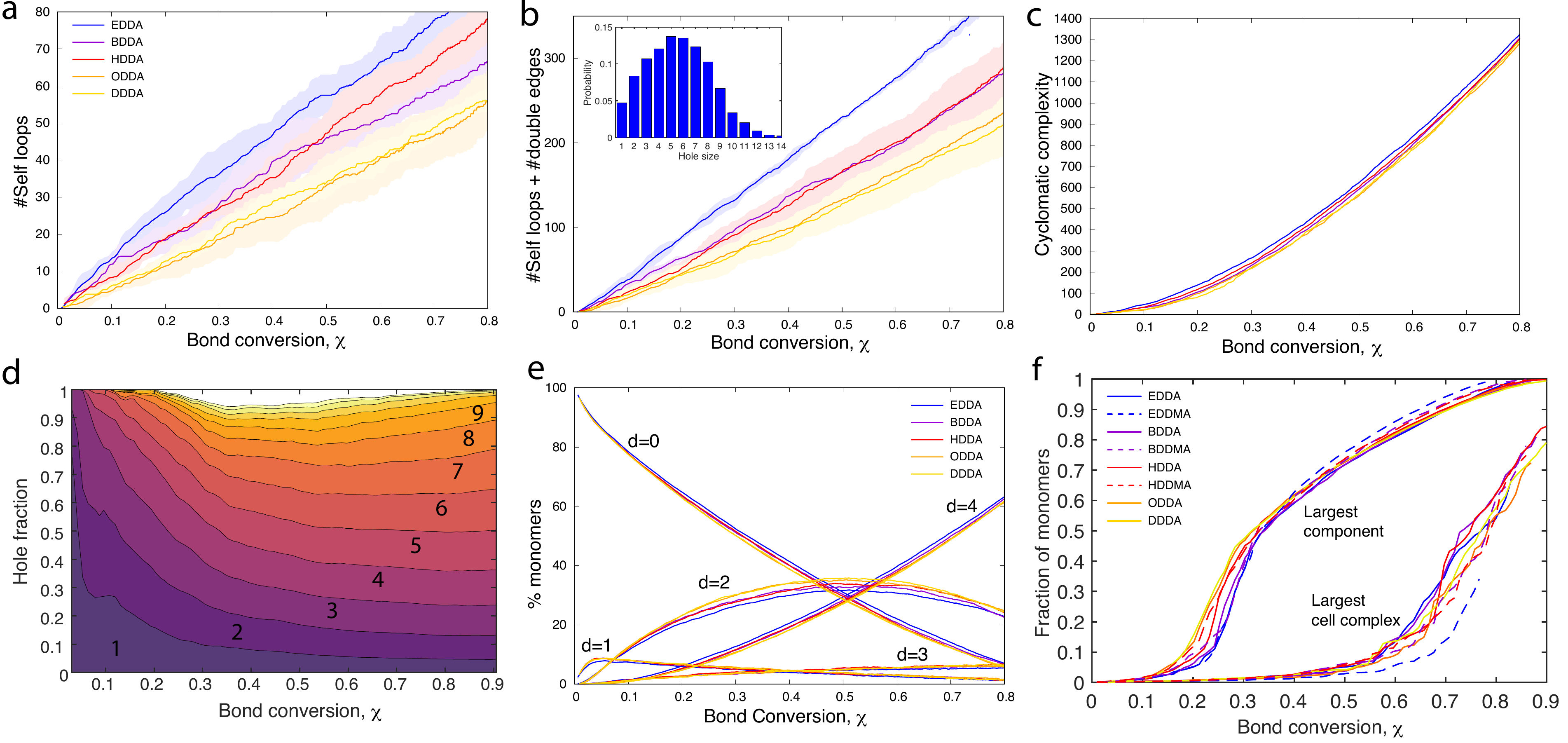}
\caption{{\bf  Graph-theoretical and topological properties of the polymer network.}
({\bf a}) Number of nodes with one self-loop.
({\bf b}) Number of nodes with either a double edge or a self loop; the inset shows an averaged distribution of hole sizes at $\chi=0.8$.
({\bf c}) Universal collapse of the overall cyclomatic complexity.
({\bf d})  The fraction of holes of indicated size as a function of bond conversion. All networks have 2000 nodes. \SIHoles~reports hole statistics for each monomer separately.
({\bf e}) Evolution of the degree distribution.
Fractions of nodes with $d$ neighbours show universal dependence for all linker lengths.
({\bf f}) Universal collapse of the sizes of the largest component and cell complex in different systems.
 The largest connected components traverses percolation transitions between
 $\chi=0.15$ and $\chi=0.25$ depending on the monomer type. The sizes of the largest connected cell complex feature a second percolation transition around $\chi=0.7$.
\SILargestComponent~reports the confidence intervals.
}
\label{fig:combi2}
\end{figure*}

Our first finding is that \emph{self loops} form frequently, as caused by a radical reacting with the pending vinyl group of the same monomer. 
Each self-loop decreases the maximum degree of one monomer from 4 to 2, and therefore self loops are of general interest in reaction kinetics and material science, where they are often referred to as topological defects  \cite{luo2020effect,sirk2020growth,Olsen2017,Olsen2016a,Olsen2016}.  
In a similar fashion, a \emph{double edge} occurs when a pair of monomers reacts twice, thus becoming connected with two bonds, which limits their external number of neighbours to 4 per pair, instead of 6. As Figure \ref{fig:combi2}a suggests, a shorter linker between the functional groups promotes formation of self loops and a very similar trend is observed for double edges, see Figure \ref{fig:combi2}b. This observation holds with one exception of HDDA. The \emph{ab initio} and classical simulations presented in Supplementary Figures S3~and~S13 suggest that this exception might be related to a coiled metastable state of HDDA  -- the end-to-end distance being close to the reaction cutoff radius. Moreover, the ability to form triangles  -- cycles of size three -- is also enhanced by short linkers, see \SITriangles, again with the BDDA-HDDA pair being swapped in the order of the trend. 

As shown in \SICycloAllMethyl~ the effect of the methyl substituent is less clear: For EDDMA, the methyl substitution induces more loops, but for the other two linker lengths (BDDMA and HDDMA) the results are inconclusive. The end-to-end distance probability distribution of the pure monomers, shown in \SIEndToEndHistogram, suggests that the larger number of self loops in EDDMA could also be caused by a metastable coiled configuration of this molecule.

Such a strong influence of monomer types on forming cyclic structures is characteristic only to the smallest scale.
To demonstrate this, we compute the maximum number of edges one can remove before the network becomes composed of tree components -- the cyclomatic complexity $r$, which can be calculated from the relationship $r=M-N+C$ with $M$ being the total number of edges, $N$ -- number of nodes, and $C$ -- number of connected components in the network.
 Clearly, self loops, double edges, and triangles increase cyclomatic complexity.
 As shown in Figure~\ref{fig:combi2}c, all monomers lead to a comparable cyclomatic complexity, where about 10\% of this quantity is explained by the self loops and double edges alone.
 In order to investigate higher order cyclic structures, we must differentiate between a \emph{cycle}, which is a closed path in a graph, and a \emph{hole} -- a cycle for which any subpath is also the shortest path connecting its own endpoints.

Bonds that are not self loops, connect different nodes together and thus extend the network in its size. By computing the degree distribution, that is fractions of nodes having different numbers of neighbours, we again observe a universal behaviour across monomer types, shown in Figure~\ref{fig:combi2}e. The dynamics seen in Figure~\ref{fig:combi2}e is different from the binomial degree distribution, which is characteristic to the standard bond percolation  \cite{Kryven2016,kryven2018,schamboeck2019,stanley1983}, and resembles the chain growth polymerisation  \cite{schamboeck2020}.
The degrees of monomers with methyl substituents are analysed in \SIDegreeDistribution, showing that the presence/absence of the methyl group does not have a significant effect on the degree distribution.
This is in strong contrast to the observed differences in polymerisation properties for different monomer types when studied as a function of time, for example as in Figure~\ref{fig:Figure2}a, where such differences are mainly attributed to diffusion limited nature of the reaction. Figure~\ref{fig:combi2}e gives reason to suggest that the diffusion effects does not strongly influence the density of the network when viewed as a function of \emph{bond conversion}. 

 Self loops and small cycles are in competition with forming bonds that expand the network size.  Although formation of small cycles does not significantly affect the degree distribution, they delay the onset of \emph{gelation}, that is the moment in time when an extensive connected component emerges.
Note that the mean-field theories predict the gel point exclusively on the basis of the degree distribution \cite{schamboeck2019,schamboeck2020a,kryven2018,Kryven2016,kryven2019bond}, which unlike self loops and small cycles, was shown to be insensitive to the monomer type.
The onset of gelation is identified by monitoring the fraction of nodes in the largest connected component.
As shown in Figure~\ref{fig:combi2}f, the gelation occurs later in bond conversion for shorter linkers. This trend is attributed to a large number of self loops that act as `defects' -- they increase the bond conversion without contributing to the expansion of the network. A similar conclusion about the effect of cyclization on the gel point was reported by Elliott \emph{et al.}  \cite{Elliott2001,Elliott1999}, where the gel point of TeGDMA was observed to be higher than the gel point of BisGDMA (bisphenol A-glycidyl dimethacrylate), with TeGDMA exhibiting three-folded more cyclization than BisGDMA due to the differences in flexibility of the monomers.

\begin{figure}
\centering
\includegraphics[width=\columnwidth]{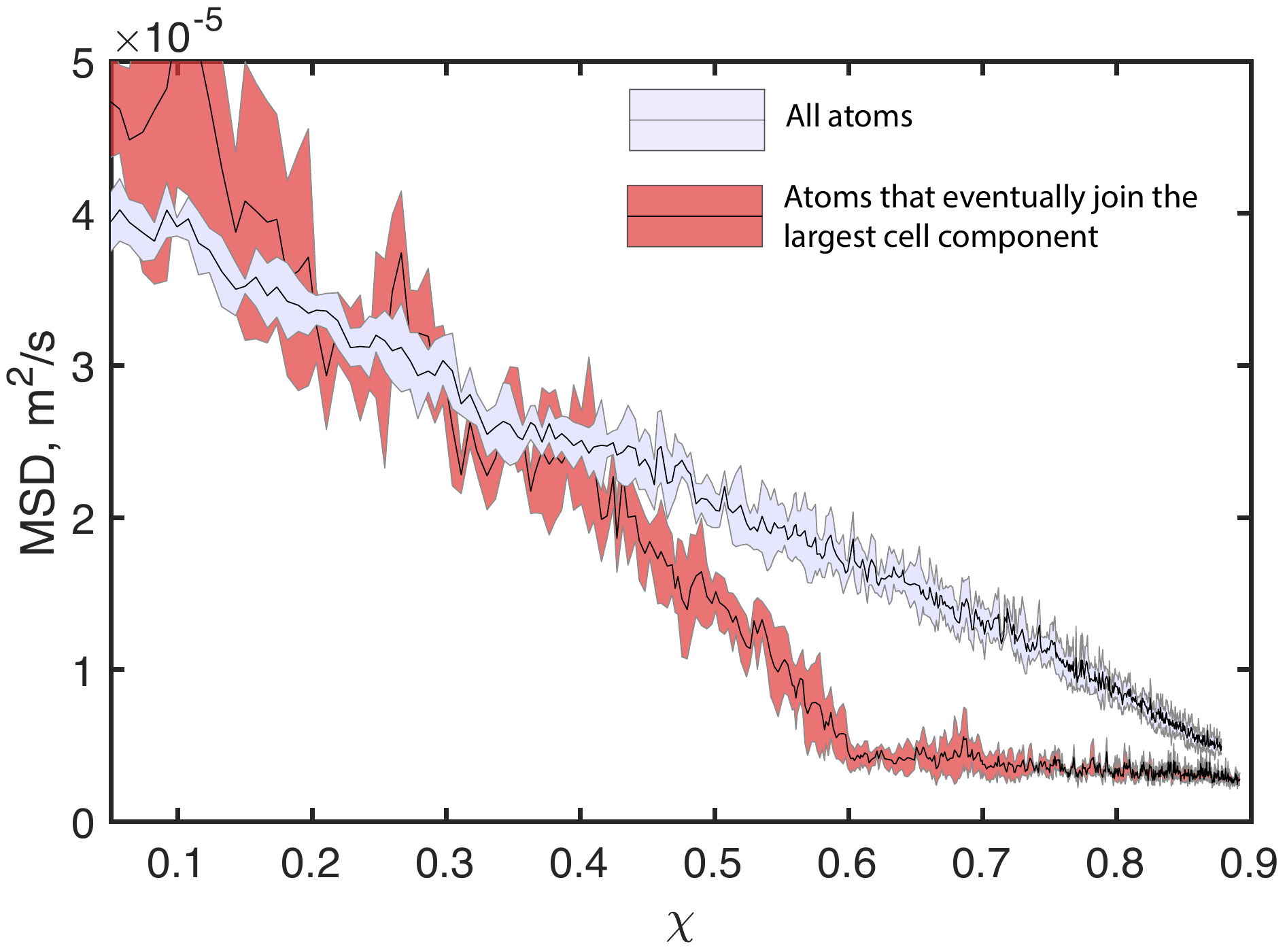}
\caption{{\bf  Mean square displacement (MSD) of atoms in the EDDA system during curing.}
The figure demonstrates a separation of the network into two phases with respect to thermal fluctuations. On average, the MSD of all atoms decreases with curing, however, the atoms that eventually join the largest connected component in the cell complex (identified at $\chi=0.6$) reach the terminal MSD  earlier than the average atom in the system.
}
\label{fig:MSD}
\end{figure}

Since the gelation starts rather early, around $\chi=0.2$, most of the network evolution takes place with the giant component being present. Moreover, around $\chi=0.4$, the whole network predominantly consists of \emph{one} connected component, plus isolated nodes. This behaviour is illustrated with network snapshots in Figure~\ref{fig:simplex}a.
Surprisingly, the connected network undergoes another structural transition, long after the onset of gelation.
We demonstrate this by analysing the \emph{cell complex}  \cite{battiston2020,bianconi2019,kryven2019} which builds upon the idea of a topological hole.
 Our construction of the cell complex can be thought of as a higher order network in which the nodes are identified with the holes in the original polymer network, and two cell-nodes are connected if the corresponding holes share vertices  \cite{iacopini2019}, see Figure~\ref{fig:simplex}b.
 This definition allows us to identify connected components in the cell complex and calculate their size.
  The sizes of the largest component in the cell complex, see Figure~\ref{fig:combi2}f, suggest that such a construction \emph{also} undergoes a percolation  transition, around $\chi=0.7$, when a percolating cell component emerges. By $\chi=0.8-0.9$, the giant cell complex expands further and incorporates the majority of the nodes.
The structure and development of the cell complex resembles a tree-like construction with a small number of higher order cycles, which can be compared to the early stages of the network formation. There is a large difference between a monomer-node in the original network and the hole-node in the cell complex -- because small holes limit the uniaxial rotation of the constituting monomers -- the latter has less degrees of freedom. A rapid change in the degrees of freedom that occurs in the physical network at the second transition, see Figure \ref{fig:MSD}, may provide a possible reason behind a sudden nonlinearity in $T_g(\chi)$ observed in Figure~\ref{fig:Figure2}c for all monomer systems.  The location of the second transition is affected by sizes of small topological holes,  which, in turn, are  influenced by monomer conformation dynamics. This mechanism allows monomers of different types to affect $T_g$, albeit only at late bond conversions after the onset of the second transition.

\section{Discussion and conclusions}
 In this work we have re-evaluated the causal link between the non-functional form of the monomer precursor and the physical properties of the resulting polymer material. To this end, we studied diacrylate monomers with different linker length and a present/absent bulky methyl group to show that in the causal chain between monomer structure and the emergent physical properties, the network topology plays a role of a unavoidable intermediary. 
 
Monomer preference to occupy certain coiled states is central to such a mechanism as it affects formation of topological holes.
 We do not entirely confirm the random-coil trend  \cite{flory1979}, longer linker -- less loops.  For instance, despite being longer than BDDA, the HDDA monomer has a stronger preference to form loops, as was shown by classical and also confirmed by the \emph{ab initio} metastability simulations. 

 Although the degree distribution is only weakly affected by the monomer type, the way the edges are arranged in the network is strongly affected as can be seen from the size distribution of topological holes. These differences affect the number of edges it takes to reach the gelation -- the moment when an extensive connected component appears and transforms formerly viscous liquid material into an elastic solid. 
We observed such a transition in two ways: topologically -- by monitoring the largest connected component in the network, and physically -- by detecting divergence of the tensile modulus from zero. Even though longer linkers were shown to slightly accelerate gelation,  the tensile modulus was found to be smaller for monomers with longer linkers. 

We showed that  although $T_g$ is not affected by gelation, there exists a novel higher-order percolation transition that drastically ramps up the $T_g$ at late bond conversions.
This  percolation transition marks the appearance of an extensive \emph{cell complex} that spans the network and therefore constrains its mobility.
In a similar way as gelation marks the onset of elastic behaviour in the polymer,  the higher order percolation transition makes the material  more brittle, as was shown by the $T_g$ tests.  Since $T_g$  suddenly becomes sensitive to monomer type at late stages, we postulate that one should be capable of influencing glass transition temperature by manipulating non-functional parts a monomer, \emph{e.g.} by changing linker length or adding non-reactive groups.
A parallel can be made here with another type of higher order structures constraining mobility of physical networks  -- a $k$-core -- that is  related to the onset of jamming in granular gels  \cite{morone2019}.

Even though we observe strong differences in the polymerisation kinetics  as a function of \emph{time} for different monomers, the topological and physical properties are surprisingly similar when reported as a function of \emph{conversion}.  Universal behaviour was seen in almost identical degree distributions, numbers of large cycles, and sizes of components far from the percolation transition. Both $T_g$ and Young's modulus feature linear dependence on the bond conversion with a universal slope at early to intermediate stages of polymerisation. 
Therefore, depending on what quantities of interest one aims to predict, coarse-grained and macroscopic models may be well  justified, for example, when used in the partial differential equations absorbing MD generated data  \cite{park2020multiscale,alame2019relative}.

\section{Methods}

\subsection*{\label{ssec:model_set_up}MD model and system set-up}
The monomers were modelled at the united-atom (UA) level with the TraPPE-UA force field by Maerzke \emph{et al.} \cite{Maerzke2009} developed for acrylates. The covalent bonds were initially represented by a harmonic potential as opposed to rigid bonds, which is necessary for the crosslinking simulations. See Tables 1-4 in the Supportive Information in reference \cite{TorresKnoop2018} for the functional form of the force field.

The initial systems were set up by randomly packing 2000 monomers in a 3D simulation cell (initial size 200$\times$200\r{A}) with periodic boundary conditions. 
The systems were then
equilibrated in the NVT ensemble at 600 K for 10 ps using a time step of 0.1fs and further equilibrated in the NPT ensemble at 600 K for 500 ps with a time step
of 1fs. This resulted in three-dimensional periodic simulation boxes of around 100$\times$100\r{A}. 
The simulations were carried out using the Nos\'{e}-Hoover thermostat \cite{Hoover1982,Nose1984}, and, in the constant pressure cases, using the Martyna-Tuckerman barostat  \cite{TuckermanMartyna}.
From the $2000$ monomers (which amounts to, \emph{e.g.} $32 000$ HDDA atoms) units, 5\% were considered active at the beginning. 

To validate force field and study the properties of the liquid monomers at ambient conditions, such as density, self-diffusivity and the viscosity of the system, the equilibrated systems at 600\,K were cooled down using the NPT ensemble from 600\,K to 300\,K and further equilibrated for 500\,ps at 300~K. The equilibrated systems at 300\,K were then used as input for longer simulations at 300\,K in the NVT ensemble to determine the density, self-diffusivity and viscosity properties of the melts, which are in a good agreement with experimental values, see \SILiquidProperties. For the viscosity, a correlation time of 1\,ns was used and the simulations were run for 500\,ns.
Further validation of the force field was done by comparing histograms and radial distribution functions (RDFs) of HDDA to  \emph{ab-initio} based molecular dynamics as implemented in the CP2K software\cite{Kuhne}.  The \emph{ab initio} simulations were performed using the BLYP functional supplemented with D3 dispersion corrections \cite{gao2020complex,Grimme}.  
As shown in \SIDihedralAngles~and  \SIEndToEndHistogram, the histogram of the dihedral angles of HDDA in liquid state and the end-to-end distance in HDDA are well captured by the TraPPE-UA force field with a few notable differences. Additionally, the RDFs of the 1-1 and 2-2 reactive sites are also well-captured by the employed force filed in comparison to the \emph{ab initio} simulations as shown in \SIRDFClassicalvsQuantum.

\subsection*{Curing simulations}
We used a simulation protocol that accelerates the curing process by lowering activation energies and increasing temperature to 600\,K, which was earlier shown not to affect the structure of the network when considered as a function of bond conversion \cite{TorresKnoop2018}.  The covalent bonds appearing during the curing of the network were created using a reactions cutoff radius of 4\,\r{A}. The proximity of reactive sites was checked every $10$\,fs and a bond was created with probability $0.5$ for every time reactive sites within the cutoff radius. After placing a new crosslink, the whole system was equilibrated in the NPT ensemble. The activation energy, which affects the chemical reaction rate, was assumed to be equal for all monomers. Hence, to link the simulation time to the real time, one needs additionally to compensate the contribution of the chemical activation energy. The curing was simulated for up to 100 ns which is enough for all systems to reach bond conversion of 0.8 in our protocol. This corresponds to 4 weeks of computational time per trajectory. The curing simulations and the tests for physical properties were repeated four times to average out the effect of the initial condition bias.

The curing of (meth)acrylates involves the addition of radicals to the unsaturated double bonds in the vinyl groups. 
Since the initiation reaction is much more slower than radical propagation, we assume that a $5\%$ of radical sites is active from the start. In experimental conditions radicals are either continuously initiated or appear in bursts as controlled by the pulses of light irradiation.  This results in a higher radical concentration than typically used in experiments, however lowering the radical concentrations would make simulations impossibly long. For the effect on the initial amount of radicals on the network formation, see Section 4 and Figures S6 and S7 in Ref.\cite{TorresKnoop2018}. 
 Assuming the initial reactive radical sites are specified, the curing was implemented in three consecutive steps:
 (1) the addition of a bond to the unsaturated carbon that undergoes a reaction, 
 (2) the change of the carbon character from unsaturated to saturated (to prevent it from participating in future reactions), and 
 (3) the propagation/regeneration of the reactive site by changing the character of the neighbouring carbon from unsaturated to the reactive one.

\subsection*{Calculations for $T_g$}
The glass transition temperature was obtained by performing NPT simulations for a series of different temperatures. 
The systems were cooled down from 600\,K to 150\,K at the rate of $5 \cdot10^{9}$\,K s$^{-1}$. We determined $T_g$  by analysing the density as a function of temperature \cite{yang2016glass} and performing a piecewise linear fit. In order to identify $T_g$ and quantify the uncertainty that arises due to the finite system size and the thermal fluctuations we introduce the following protocol for statistical inference: First, we solved the minimisation problem 
\begin{align*}
j^* = \arg\min_{j} \sum_{i=1}^j \| r_{i,1}\|_2+\sum_{i=j+1}^n \|r_{i,2} \|_2 
\end{align*}
where 
$$r_{i,1}= \rho_i - (\beta_{j,1} T_i +\varepsilon_{j,1}),$$
$$r_{i,2}= \rho_i - (\beta_{j,2} T_i +\varepsilon_{j,2}),$$
are the residuals of least square linear fits that correspond to the first ($\rho_i,T_i,\; i=1,\dots,j$) and second ($\rho_i,T_i,\; i=j+1,\dots,n$) fragments of data points separated by index $j$; 
$\rho_i,$ $i=1,\dots,n$ are the measured  density at temperatures $T_i,$ and $\beta_{j,k},\varepsilon_{j,k},\; k=1,2$ are the coefficients of the fits. We then define the glass transition temperature as the point where the linear fits intersect:
$$
T_g = \frac{\varepsilon_{j,2} -\varepsilon_{j,1}}{\beta_{j,1}-\beta_{j,2}}.
$$
Even though $\rho_i$ are normally distributed around their fits, $T_g$ is  defined as a quotient, and therefore, is not normally nor symmetrically distributed. In fact the distribution for $T_g$ may even contain a heavy tail when the intersection angle is close to $\pi$. We thus estimate the confidence intervals for $T_g$ by using a sampling method. 
Let us denote the averages of the first and second intervals as $\langle x_i\rangle_1:=\frac{1}{j^*}\sum_{i=1}^{j^*}x_i$ and $\langle x_i\rangle_2:=\frac{1}{n-j^*+1}\sum_{i=j^*+1}^{n}x_i$. We  define
\begin{align*}
\sigma^2_1:=\langle T_i \rangle_2^2 \frac{ \langle r_{i,1}^2 \rangle_1- \langle r_{i,1}\rangle_1  ^2  }{\langle T_i^2 \rangle_1   - j^* 
\langle T_i \rangle_1^2 }+
\langle T_i \rangle_1^2 \frac{ \langle r_{i,1}^2 \rangle_2- \langle r_{i,1}\rangle_2  ^2  }{\langle T_i^2 \rangle_2   - (n-j^*+1) 
\langle T_i \rangle_2^2 }
\end{align*}
and
\begin{align*}
\sigma^2_2:=\frac{ \langle r_{i,1}^2 \rangle_1- \langle r_{i,1}\rangle_1  ^2  }{\langle T_i^2 \rangle_1   - j^* 
\langle T_i \rangle_1^2 }+
 \frac{ \langle r_{i,1}^2 \rangle_2- \langle r_{i,1}\rangle_2  ^2  }{\langle T_i^2 \rangle_2   - (n-j^*+1) 
\langle T_i \rangle_2^2 }.
\end{align*}
Let  $X_i\sim\mathcal{N}(\varepsilon_{j,2} -\varepsilon_{j,1},\sigma^2_1 )$ and  $X_i\sim\mathcal{N}(\beta_{j,1}-\beta_{j,2},\sigma^2_2)$ be $N$ independently generated samples generated from the normal distribution $\mathcal{N}(\mu, \sigma^2)$ with mean $\mu$ and variance $\sigma^2$.
Then $Z_i=\frac{X_i}{Y_i}$ are random samples from the distribution for the $T_g$ estimate.
The confidence intervals for $T_g$ are the percentiles of the empirical density function for $Z_i-\langle Z_i \rangle$. 

\subsection*{Calculation of Young's modulus}
To determine Young’s modulus, we performed uniaxial
tension loading simulations by increasing the length of the simulation cell along the loading direction at every MD step while
maintaining atmospheric pressure in the transverse directions using a barostat. Young's modulus was determined by estimating the slope of the engineering stress-strain curve in the linear region, when a sample is pulled at a constant rate in the static tensile test. 
The stress-strain curves were obtained by applying a uniaxial deformation along the $x$-axis at a constant rate and measuring the strain-stress response curves. 
The limit of the longitudinal strain was set to be 50\%, and the strain rate was varied between $10^{8}-10^{10}$s$^{-1}$ showing no statistically significant dependence of the inferred Young's modulus on the rate, see \SIStrainRate. 
The strain was determined from  $L(t) = L_0\times \exp(\text{rate}\times {\rm d}t)$, where $L(t)$ and $L_0$ are respectively the instantaneous and initial length of the simulation box in the direction of elongation. The stress was calculated from the virial expansion. The transverse directions were maintained at atmospheric pressure using the Martyna-Tuckerman barostat  \cite{TuckermanMartyna}. 
Since the test was performed under conditions of thermal fluctuations, as illustrated in Figure~\ref{fig:Figure2}d,  we used Ljung-Box Q-Test to determine if fluctuations around the linear fit for the first $j=2,3,...$ data-points are pair-correlated. Then, if the identified fragment  of the data series also passed the Kolmogorov Smirnov test for normality, the slope and its confidence interval were deduced applying the linear regression. The linear region was identified as the largest strain for which these tests were passed. The length of the linear region typically fluctuated between 0.06-0.2, with higher bond conversion being associated with shorter linear regions.
 This procedure was  repeated for different conversions $\chi$ and different monomer types, as reported in \SIYoung.  All  data points together with their confidence intervals  across the conversion range were fitted again with a straight line, as shown in Figure \ref{fig:Figure2}e. The `fuzziness' of the latter fits is due to confidence intervals of data points in the latter fits. Such secondary fits are therefore drawn from a probability distribution.
Note that in the viscous regime (before the onset of gelation), our method inferred no values for the modulus because the data violated Ljung-Box Q-Test.

\subsection*{Average resistance model for elasticity}
The approximation of the elastic modulus was obtained using the  graph resistance model:
$$E^*=C\left(\frac{1}{|\Omega_1| |\Omega_2|}\sum\limits_{i\in\Omega_1,\;j\in\Omega_2}(L^+_{i,i}+L^+_{j,j}-L^+_{i,j}-L^+_{j,i})\right)^{-1}\!\!,$$
where $C$ is a constant that represents monomer tendency to coil and (was  estimated from molecular simulations),   
$\bf L=D-A$ is the graph laplacian with $\bf D$ being the diagonal matrix with node degrees on the diagonal and
$\bf A$ the adjacency matrix of the monomer network and $\bf L^+$ indicating its Moore-Penrose inverse; $\Omega_1$ and $\Omega_2$ are the index sets for monomers that are located at the two opposite faces of the simulation box.
This estimate is based on the electrical engineering concept of effective resistance \cite{ellens2011}. Note that the method is not applicable when there is no percolating cluster between $\Omega_1$ and $\Omega_2$.

\section*{Acknowledgements}
ATK and VS acknowledge the financial support from the Netherlands Organisation for Scientific Research (NWO) via respectively project PREDAGIO and the Open Technology Program.

\section*{Author contributions}
ATK and VS -- high-performance computing and molecular dynamics,
NG -- \emph{ab initio} simulations, 
PDI -- chemical context,
IK -- research design and network topology analysis.
All authors discussed results and contributed to writing.

\section*{Additional information}
The authors report no competing interests.

\section*{Data availability}
All relevant data are available from the authors on reasonable request.


\begin{thebibliography}{10}
\expandafter\ifx\csname url\endcsname\relax
  \def\url#1{\texttt{#1}}\fi
\expandafter\ifx\csname urlprefix\endcsname\relax\def\urlprefix{URL }\fi
\providecommand{\bibinfo}[2]{#2}
\providecommand{\eprint}[2][]{\url{#2}}

\bibitem{de1976relation}
\bibinfo{author}{De~Gennes, P.-G.}
\newblock \bibinfo{title}{On a relation between percolation theory and the
  elasticity of gels}.
\newblock \emph{\bibinfo{journal}{Journal de Physique Lettres}}
  \textbf{\bibinfo{volume}{37}}, \bibinfo{pages}{1--2} (\bibinfo{year}{1976}).

\bibitem{stockmayer1943theory}
\bibinfo{author}{Stockmayer, W.~H.}
\newblock \bibinfo{title}{Theory of molecular size distribution and gel
  formation in branched-chain polymers}.
\newblock \emph{\bibinfo{journal}{The Journal of chemical physics}}
  \textbf{\bibinfo{volume}{11}}, \bibinfo{pages}{45--55}
  (\bibinfo{year}{1943}).

\bibitem{flory1969statistical}
\bibinfo{author}{Flory, P.~J.} \& \bibinfo{author}{Volkenstein, M.}
\newblock \bibinfo{title}{Statistical mechanics of chain molecules}
  (\bibinfo{year}{1969}).

\bibitem{ziff1980kinetics}
\bibinfo{author}{Ziff, R.~M.} \& \bibinfo{author}{Stell, G.}
\newblock \bibinfo{title}{Kinetics of polymer gelation}.
\newblock \emph{\bibinfo{journal}{The Journal of Chemical Physics}}
  \textbf{\bibinfo{volume}{73}}, \bibinfo{pages}{3492--3499}
  (\bibinfo{year}{1980}).

\bibitem{Kryven2016}
\bibinfo{author}{Kryven, I.}
\newblock \bibinfo{title}{Emergence of the giant weak component in directed
  random graphs with arbitrary degree distributions}.
\newblock \emph{\bibinfo{journal}{Phys. Rev. E}} \textbf{\bibinfo{volume}{94}},
  \bibinfo{pages}{012315} (\bibinfo{year}{2016}).

\bibitem{Kryven2017}
\bibinfo{author}{Kryven, I.}
\newblock \bibinfo{title}{General expression for the component size
  distribution in infinite configuration networks}.
\newblock \emph{\bibinfo{journal}{Phys. Rev. E}} \textbf{\bibinfo{volume}{95}},
  \bibinfo{pages}{052303} (\bibinfo{year}{2017}).

\bibitem{kryven2018}
\bibinfo{author}{Kryven, I.}
\newblock \bibinfo{title}{Analytic results on the polymerisation random graph
  model}.
\newblock \emph{\bibinfo{journal}{Journal of Mathematical Chemistry}}
  \textbf{\bibinfo{volume}{56}}, \bibinfo{pages}{140--157}
  (\bibinfo{year}{2018}).

\bibitem{schamboeck2019}
\bibinfo{author}{Schamboeck, V.}, \bibinfo{author}{Iedema, P.~D.} \&
  \bibinfo{author}{Kryven, I.}
\newblock \bibinfo{title}{Dynamic networks that drive the process of
  irreversible step-growth polymerization}.
\newblock \emph{\bibinfo{journal}{Scientific reports}}
  \textbf{\bibinfo{volume}{9}}, \bibinfo{pages}{1--18} (\bibinfo{year}{2019}).

\bibitem{rudyak2019thermoset}
\bibinfo{author}{Rudyak, V.}, \bibinfo{author}{Efimova, E.},
  \bibinfo{author}{Guseva, D.} \& \bibinfo{author}{Chertovich, A.}
\newblock \bibinfo{title}{Thermoset polymer matrix structure and properties:
  Coarse-grained simulations}.
\newblock \emph{\bibinfo{journal}{Polymers}} \textbf{\bibinfo{volume}{11}},
  \bibinfo{pages}{36} (\bibinfo{year}{2019}).

\bibitem{klahn2019effect}
\bibinfo{author}{Kl{\"a}hn, M.} \emph{et~al.}
\newblock \bibinfo{title}{Effect of external and internal plasticization on the
  glass transition temperature of (meth) acrylate polymers studied with
  molecular dynamics simulations and calorimetry}.
\newblock \emph{\bibinfo{journal}{Polymer}} \textbf{\bibinfo{volume}{179}},
  \bibinfo{pages}{121635} (\bibinfo{year}{2019}).

\bibitem{huang2019effects}
\bibinfo{author}{Huang, M.} \& \bibinfo{author}{Abrams, C.}
\newblock \bibinfo{title}{Effects of reactivity ratios on network topology and
  thermomechanical properties in vinyl ester/styrene thermosets: Molecular
  dynamics simulations}.
\newblock \emph{\bibinfo{journal}{Macromolecular Theory and Simulations}}
  \bibinfo{pages}{1900030} (\bibinfo{year}{2019}).

\bibitem{demir2019versatile}
\bibinfo{author}{Demir, B.} \& \bibinfo{author}{Walsh, T.~R.}
\newblock \bibinfo{title}{A versatile computational procedure for chain-growth
  polymerization using molecular dynamics simulations}.
\newblock \emph{\bibinfo{journal}{ACS Applied Polymer Materials}}
  (\bibinfo{year}{2019}).

\bibitem{jung2019free}
\bibinfo{author}{Jung, J.}, \bibinfo{author}{Park, C.} \& \bibinfo{author}{Yun,
  G.~J.}
\newblock \bibinfo{title}{Free radical polymerization simulation and molecular
  entanglement effect on large deformation behavior}.
\newblock \emph{\bibinfo{journal}{European Polymer Journal}}
  \textbf{\bibinfo{volume}{114}}, \bibinfo{pages}{223--233}
  (\bibinfo{year}{2019}).

\bibitem{Gissinger2020}
\bibinfo{author}{Gissinger, J.~R.}, \bibinfo{author}{Jensen, B.~D.} \&
  \bibinfo{author}{Wise, K.~E.}
\newblock \bibinfo{title}{Reacter: A heuristic method for reactive molecular
  dynamics}.
\newblock \emph{\bibinfo{journal}{Macromolecules}}
  \textbf{\bibinfo{volume}{53}}, \bibinfo{pages}{9953--9961}
  (\bibinfo{year}{2020}).

\bibitem{TorresKnoop2018}
\bibinfo{author}{Torres-Knoop, A.}, \bibinfo{author}{Kryven, I.},
  \bibinfo{author}{Schamboeck, V.} \& \bibinfo{author}{Iedema, P.~D.}
\newblock \bibinfo{title}{Modeling the free-radical polymerization of
  hexanediol diacrylate (hdda): a molecular dynamics and graph theory
  approach}.
\newblock \emph{\bibinfo{journal}{Soft Matter}} \textbf{\bibinfo{volume}{14}},
  \bibinfo{pages}{3404--3414} (\bibinfo{year}{2018}).

\bibitem{wang2020room}
\bibinfo{author}{Wang, H.} \emph{et~al.}
\newblock \bibinfo{title}{Room-temperature autonomous self-healing glassy
  polymers with hyperbranched structure}.
\newblock \emph{\bibinfo{journal}{Proceedings of the National Academy of
  Sciences}} \textbf{\bibinfo{volume}{117}}, \bibinfo{pages}{11299--11305}
  (\bibinfo{year}{2020}).

\bibitem{bio1}
\bibinfo{author}{Lu, H.}, \bibinfo{author}{Stansbury, J.~W.},
  \bibinfo{author}{Nie, J.}, \bibinfo{author}{Berchtold, K.~A.} \&
  \bibinfo{author}{Bowman, C.~N.}
\newblock \bibinfo{title}{Development of highly reactive mono-(meth)acrylates
  as reactive diluents for dimethacrylate-based dental resin systems}.
\newblock \emph{\bibinfo{journal}{Biomaterials}} \textbf{\bibinfo{volume}{26}},
  \bibinfo{pages}{1329--1336} (\bibinfo{year}{2005}).

\bibitem{lito}
\bibinfo{author}{Tran, K.~T.} \& \bibinfo{author}{Nguyen, T.~D.}
\newblock \bibinfo{title}{Lithography-based methods to manufacture biomaterials
  at small scale}.
\newblock \emph{\bibinfo{journal}{Journal of Science: Advanced Materials and
  Devices}} \textbf{\bibinfo{volume}{2}}, \bibinfo{pages}{1--14}
  (\bibinfo{year}{2017}).

\bibitem{bio2}
\bibinfo{author}{Moszner, N.} \& \bibinfo{author}{Salz, U.}
\newblock \bibinfo{title}{New developments of polymeric dental composites}.
\newblock \emph{\bibinfo{journal}{Prog. Polym. Sci.}}
  \textbf{\bibinfo{volume}{26}}, \bibinfo{pages}{535--576}
  (\bibinfo{year}{2001}).

\bibitem{coating}
\bibinfo{author}{Decker, C.}, \bibinfo{author}{Viet, T. N.~T.} \&
  \bibinfo{author}{Decker, D.}
\newblock \bibinfo{title}{Uv-radiation curing of acrylate/epoxide systems}.
\newblock \emph{\bibinfo{journal}{Polymer}} \textbf{\bibinfo{volume}{42}},
  \bibinfo{pages}{5531--5541} (\bibinfo{year}{2001}).

\bibitem{coatings2}
\bibinfo{author}{Yao, B.} \emph{et~al.}
\newblock \bibinfo{title}{Synthesis of acrylate-based uv/thermal dual-cure
  coatings for antifogging}.
\newblock \emph{\bibinfo{journal}{J. Coat. Technol. Res.}}
  \textbf{\bibinfo{volume}{15}}, \bibinfo{pages}{149--158}
  (\bibinfo{year}{2018}).

\bibitem{bio3}
\bibinfo{author}{Anseth, K.~S.} \& \bibinfo{author}{Burdick, J.~A.}
\newblock \bibinfo{title}{New directions in photopolymerizable biomaterials}.
\newblock \emph{\bibinfo{journal}{MRS Bull.}} \textbf{\bibinfo{volume}{27}},
  \bibinfo{pages}{130--136} (\bibinfo{year}{2002}).

\bibitem{bio4}
\bibinfo{author}{Fisher, J.~P.}, \bibinfo{author}{Dean, D.},
  \bibinfo{author}{Engle, P.~S.} \& \bibinfo{author}{Mikos, A.~G.}
\newblock \bibinfo{title}{Photoinitiated polymerization of biomaterials}.
\newblock \emph{\bibinfo{journal}{Annu. Rev. Mater. Res.}}
  \textbf{\bibinfo{volume}{31}}, \bibinfo{pages}{171--181}
  (\bibinfo{year}{2001}).

\bibitem{vanSteenberge2019}
\bibinfo{author}{Van~Steenberge, P.~H.} \emph{et~al.}
\newblock \bibinfo{title}{Visualization and design of the functional group
  distribution during statistical copolymerization}.
\newblock \emph{\bibinfo{journal}{Nature communications}}
  \textbf{\bibinfo{volume}{10}}, \bibinfo{pages}{1--14} (\bibinfo{year}{2019}).

\bibitem{gao2020complex}
\bibinfo{author}{Gao, Y.} \emph{et~al.}
\newblock \bibinfo{title}{Complex polymer architectures through free-radical
  polymerization of multivinyl monomers}.
\newblock \emph{\bibinfo{journal}{Nature Reviews Chemistry}}
  \bibinfo{pages}{1--19} (\bibinfo{year}{2020}).

\bibitem{Kurdikar1994}
\bibinfo{author}{Kurdikar, D.~L.} \& \bibinfo{author}{Peppas, N.~A.}
\newblock \bibinfo{title}{A kinetic study of diacrylate photopolymerizations}.
\newblock \emph{\bibinfo{journal}{Polymer}} \textbf{\bibinfo{volume}{35}},
  \bibinfo{pages}{1004--1011} (\bibinfo{year}{1994}).

\bibitem{Buback2009}
\bibinfo{author}{Buback, M.}
\newblock \bibinfo{title}{Fundamentals of free-radical polymerization
  propagation kinetics in radical polymerization studied via pulsed laser
  techniques}.
\newblock \emph{\bibinfo{journal}{Macromol Symp.}} \bibinfo{pages}{275--276}
  (\bibinfo{year}{2009}).

\bibitem{Wen1997}
\bibinfo{author}{Wen, M.} \emph{et~al.}
\newblock \bibinfo{title}{Kinetic study of free-radical polymerization of
  multifunctional acrylates adn methacrylates}.
\newblock \emph{\bibinfo{journal}{IS\&T's 50th Annual Conference}}
  \bibinfo{pages}{564} (\bibinfo{year}{1997}).

\bibitem{Yu2008}
\bibinfo{author}{Yu, X.}, \bibinfo{author}{Pfaendtner, J.} \&
  \bibinfo{author}{Broadbelt, L.~J.}
\newblock \bibinfo{title}{Ab initio study of acrylate polymerization reactions:
  Methyl methacrylate and methyl acrylate propagation}.
\newblock \emph{\bibinfo{journal}{J. Phys. Chem. A}}
  \textbf{\bibinfo{volume}{112}}, \bibinfo{pages}{6772--6782}
  (\bibinfo{year}{2008}).

\bibitem{Mavroudakis2015}
\bibinfo{author}{Mavroudakis, E.}, \bibinfo{author}{Cuccato, D.},  \&
  \bibinfo{author}{Moscatelli, D.}
\newblock \bibinfo{title}{On the use of quantum chemistry for the determination
  of propagation, copolymerization, and secondary reaction kinetics in free
  radical polymerization}.
\newblock \emph{\bibinfo{journal}{Polymers}} \textbf{\bibinfo{volume}{7}},
  \bibinfo{pages}{1789--1891} (\bibinfo{year}{2015}).

\bibitem{yang2016glass}
\bibinfo{author}{Yang, Q.}, \bibinfo{author}{Chen, X.}, \bibinfo{author}{He,
  Z.}, \bibinfo{author}{Lan, F.} \& \bibinfo{author}{Liu, H.}
\newblock \bibinfo{title}{The glass transition temperature measurements of
  polyethylene: determined by using molecular dynamic method}.
\newblock \emph{\bibinfo{journal}{Rsc Advances}} \textbf{\bibinfo{volume}{6}},
  \bibinfo{pages}{12053--12060} (\bibinfo{year}{2016}).

\bibitem{kannurpatti1998study}
\bibinfo{author}{Kannurpatti, A.~R.}, \bibinfo{author}{Anseth, J.~W.} \&
  \bibinfo{author}{Bowman, C.~N.}
\newblock \bibinfo{title}{A study of the evolution of mechanical properties and
  structural heterogeneity of polymer networks formed by photopolymerizations
  of multifunctional (meth) acrylates}.
\newblock \emph{\bibinfo{journal}{Polymer}} \textbf{\bibinfo{volume}{39}},
  \bibinfo{pages}{2507--2513} (\bibinfo{year}{1998}).

\bibitem{Bowman1990}
\bibinfo{author}{Bowman, C.~N.}, \bibinfo{author}{Carver, A.~L.},
  \bibinfo{author}{Kennett, S.~N.} \& \bibinfo{author}{Peppas, N.~A.}
\newblock \bibinfo{title}{Polymers for information storage systems iii.
  crosslinked structure of polydimethacrylate}.
\newblock \emph{\bibinfo{journal}{Polymer}} \textbf{\bibinfo{volume}{31}},
  \bibinfo{pages}{135--139} (\bibinfo{year}{1990}).

\bibitem{Kurdikar1995}
\bibinfo{author}{Kurdikar, D.} \& \bibinfo{author}{Peppas, N.~A.}
\newblock \bibinfo{title}{The volume shrinkage, thermal and sorption behaviour
  of polydiacrylates}.
\newblock \emph{\bibinfo{journal}{Polymer}} \textbf{\bibinfo{volume}{36}},
  \bibinfo{pages}{2249--2255} (\bibinfo{year}{1995}).

\bibitem{Jerolimov1994}
\bibinfo{author}{Jerolimov, V.}, \bibinfo{author}{Jagger, R.~G.} \&
  \bibinfo{author}{Millward, P.~J.}
\newblock \bibinfo{title}{Effect of cross-linking chain lenght on glass
  transition of a dough-moulded poly(methylmethacrylate) resins}.
\newblock \emph{\bibinfo{journal}{Atca Stomatol. Crota.}}
  \textbf{\bibinfo{volume}{28}}, \bibinfo{pages}{3--9} (\bibinfo{year}{1994}).

\bibitem{cook1990influence}
\bibinfo{author}{Cook, W.~D.} \& \bibinfo{author}{Moopnar, M.}
\newblock \bibinfo{title}{Influence of chemical structure on the fracture
  behaviour of dimethacrylate composite resins}.
\newblock \emph{\bibinfo{journal}{Biomaterials}} \textbf{\bibinfo{volume}{11}},
  \bibinfo{pages}{272--276} (\bibinfo{year}{1990}).

\bibitem{davis1988properties}
\bibinfo{author}{Davis, T.~P.}, \bibinfo{author}{Huglin, M.~B.} \&
  \bibinfo{author}{Yip, D.~C.}
\newblock \bibinfo{title}{Properties of poly (n-vinyl-2-pyrrolidone) hydrogels
  crosslinked with ethyleneglycol dimethacrylate}.
\newblock \emph{\bibinfo{journal}{Polymer}} \textbf{\bibinfo{volume}{29}},
  \bibinfo{pages}{701--706} (\bibinfo{year}{1988}).

\bibitem{luo2020effect}
\bibinfo{author}{Luo, K.}, \bibinfo{author}{Wangari, C.},
  \bibinfo{author}{Subhash, G.} \& \bibinfo{author}{Spearot, D.~E.}
\newblock \bibinfo{title}{Effect of loop defects on the high strain rate
  behavior of pegda hydrogels: A molecular dynamics study}.
\newblock \emph{\bibinfo{journal}{The Journal of Physical Chemistry B}}
  \textbf{\bibinfo{volume}{124}}, \bibinfo{pages}{2029--2039}
  (\bibinfo{year}{2020}).

\bibitem{sirk2020growth}
\bibinfo{author}{Sirk, T.~W.}
\newblock \bibinfo{title}{Growth and arrest of topological cycles in small
  physical networks}.
\newblock \emph{\bibinfo{journal}{Proceedings of the National Academy of
  Sciences}}  (\bibinfo{year}{2020}).

\bibitem{Olsen2017}
\bibinfo{author}{Gu, Y.} \emph{et~al.}
\newblock \bibinfo{title}{Semibatch monomer addition as a general method to
  tune and enhance the mechanics of polymer networks via loop-defect control}.
\newblock \emph{\bibinfo{journal}{Proceedings of the National Academy of
  Sciences}} \textbf{\bibinfo{volume}{114}}, \bibinfo{pages}{4875--4880}
  (\bibinfo{year}{2017}).

\bibitem{Olsen2016a}
\bibinfo{author}{Zhong, M.}, \bibinfo{author}{Wang, R.},
  \bibinfo{author}{Kawamoto, K.}, \bibinfo{author}{Olsen, B.~D.} \&
  \bibinfo{author}{Johnson, J.~A.}
\newblock \bibinfo{title}{Quantifying the impact of molecular defects on
  polymer network elasticity} \textbf{\bibinfo{volume}{353}},
  \bibinfo{pages}{1264--1268} (\bibinfo{year}{2016}).

\bibitem{Olsen2016}
\bibinfo{author}{Wang, R.}, \bibinfo{author}{Alexander-Katz, A.},
  \bibinfo{author}{Johnson, J.~A.} \& \bibinfo{author}{Olsen, B.~D.}
\newblock \bibinfo{title}{Universal cyclic topology in polymer networks}.
\newblock \emph{\bibinfo{journal}{Phys. Rev. Lett.}}
  \textbf{\bibinfo{volume}{116}}, \bibinfo{pages}{188302}
  (\bibinfo{year}{2016}).

\bibitem{stanley1983}
\bibinfo{author}{Stanley, H.~E.}, \bibinfo{author}{Blumberg, R.~L.} \&
  \bibinfo{author}{Geiger, A.}
\newblock \bibinfo{title}{Gelation models of hydrogen bond networks in liquid
  water}.
\newblock \emph{\bibinfo{journal}{Physical Review B}}
  \textbf{\bibinfo{volume}{28}}, \bibinfo{pages}{1626} (\bibinfo{year}{1983}).

\bibitem{schamboeck2020}
\bibinfo{author}{Schamboeck, V.}, \bibinfo{author}{Iedema, P.~D.} \&
  \bibinfo{author}{Kryven, I.}
\newblock \bibinfo{title}{Coloured random graphs explain the structure and
  dynamics of cross-linked polymer networks}.
\newblock \emph{\bibinfo{journal}{Scientific Reports}}
  \textbf{\bibinfo{volume}{10}}, \bibinfo{pages}{1--18} (\bibinfo{year}{2020}).

\bibitem{schamboeck2020a}
\bibinfo{author}{Schamboeck, V.}, \bibinfo{author}{Kryven, I.} \&
  \bibinfo{author}{Iedema, P.~D.}
\newblock \bibinfo{title}{Effect of volume growth on the percolation threshold
  in random directed acyclic graphs with a given degree distribution}.
\newblock \emph{\bibinfo{journal}{Physical Review E}}
  \textbf{\bibinfo{volume}{101}}, \bibinfo{pages}{012303}
  (\bibinfo{year}{2020}).

\bibitem{kryven2019bond}
\bibinfo{author}{Kryven, I.}
\newblock \bibinfo{title}{Bond percolation in coloured and multiplex networks}.
\newblock \emph{\bibinfo{journal}{Nature communications}}
  \textbf{\bibinfo{volume}{10}}, \bibinfo{pages}{1--16} (\bibinfo{year}{2019}).

\bibitem{Elliott2001}
\bibinfo{author}{Elliott, J.~E.}, \bibinfo{author}{Lovell, L.~G.} \&
  \bibinfo{author}{Bowman, C.~N.}
\newblock \bibinfo{title}{Primary cyclization in the polymerization of bis-gma
  and tegdma: a modeling approach to understanding the cure of dental resins}.
\newblock \emph{\bibinfo{journal}{Dent Mater.}} \textbf{\bibinfo{volume}{17}},
  \bibinfo{pages}{221–9} (\bibinfo{year}{2001}).

\bibitem{Elliott1999}
\bibinfo{author}{Elliott, J.~E.} \& \bibinfo{author}{Bowman, C.~N.}
\newblock \bibinfo{title}{Kinetics of primary cyclyzation reactions in
  cross-linked polymers: An analytical and numerical approach to heterogeneity
  in networks}.
\newblock \emph{\bibinfo{journal}{Macromolecules}}
  \textbf{\bibinfo{volume}{32}}, \bibinfo{pages}{8621--8628}
  (\bibinfo{year}{1999}).

\bibitem{battiston2020}
\bibinfo{author}{Battiston, F.} \emph{et~al.}
\newblock \bibinfo{title}{Networks beyond pairwise interactions: structure and
  dynamics}.
\newblock \emph{\bibinfo{journal}{Physics Reports}}  (\bibinfo{year}{2020}).

\bibitem{bianconi2019}
\bibinfo{author}{Bianconi, G.}, \bibinfo{author}{Kryven, I.} \&
  \bibinfo{author}{Ziff, R.~M.}
\newblock \bibinfo{title}{Percolation on branching simplicial and cell
  complexes and its relation to interdependent percolation}.
\newblock \emph{\bibinfo{journal}{Physical Review E}}
  \textbf{\bibinfo{volume}{100}}, \bibinfo{pages}{062311}
  (\bibinfo{year}{2019}).

\bibitem{kryven2019}
\bibinfo{author}{Kryven, I.}, \bibinfo{author}{Ziff, R.~M.} \&
  \bibinfo{author}{Bianconi, G.}
\newblock \bibinfo{title}{Renormalization group for link percolation on planar
  hyperbolic manifolds}.
\newblock \emph{\bibinfo{journal}{Physical Review E}}
  \textbf{\bibinfo{volume}{100}}, \bibinfo{pages}{022306}
  (\bibinfo{year}{2019}).

\bibitem{iacopini2019}
\bibinfo{author}{Iacopini, I.}, \bibinfo{author}{Petri, G.},
  \bibinfo{author}{Barrat, A.} \& \bibinfo{author}{Latora, V.}
\newblock \bibinfo{title}{Simplicial models of social contagion}.
\newblock \emph{\bibinfo{journal}{Nature communications}}
  \textbf{\bibinfo{volume}{10}}, \bibinfo{pages}{1--9} (\bibinfo{year}{2019}).

\bibitem{flory1979}
\bibinfo{author}{Flory, P.~J.}
\newblock \bibinfo{title}{Molecular theory of rubber elasticity}.
\newblock \emph{\bibinfo{journal}{Polymer}} \textbf{\bibinfo{volume}{20}},
  \bibinfo{pages}{1317--1320} (\bibinfo{year}{1979}).

\bibitem{morone2019}
\bibinfo{author}{Morone, F.}, \bibinfo{author}{Burleson-Lesser, K.},
  \bibinfo{author}{Vinutha, H.}, \bibinfo{author}{Sastry, S.} \&
  \bibinfo{author}{Makse, H.~A.}
\newblock \bibinfo{title}{The jamming transition is a k-core percolation
  transition}.
\newblock \emph{\bibinfo{journal}{Physica A: Statistical Mechanics and its
  Applications}} \textbf{\bibinfo{volume}{516}}, \bibinfo{pages}{172--177}
  (\bibinfo{year}{2019}).

\bibitem{park2020multiscale}
\bibinfo{author}{Park, C.}, \bibinfo{author}{Jung, J.} \& \bibinfo{author}{Yun,
  G.~J.}
\newblock \bibinfo{title}{Multiscale micromorphic theory compatible with md
  simulations in both time-scale and length-scale}.
\newblock \emph{\bibinfo{journal}{International Journal of Plasticity}}
  \bibinfo{pages}{102680} (\bibinfo{year}{2020}).

\bibitem{alame2019relative}
\bibinfo{author}{Alam{\'e}, G.} \& \bibinfo{author}{Brassart, L.}
\newblock \bibinfo{title}{Relative contributions of chain density and topology
  to the elasticity of two-dimensional polymer networks}.
\newblock \emph{\bibinfo{journal}{Soft matter}} \textbf{\bibinfo{volume}{15}},
  \bibinfo{pages}{5703--5713} (\bibinfo{year}{2019}).

\bibitem{Maerzke2009}
\bibinfo{author}{Maerzke, K.~A.}, \bibinfo{author}{Schultz, N.~E.},
  \bibinfo{author}{Ross, R.~B.} \& \bibinfo{author}{Siepmann, J.~I.}
\newblock \bibinfo{title}{Trappe-ua force field for acrylates and monte carlo
  simulations for their mixtures with alkanes and alcohols}.
\newblock \emph{\bibinfo{journal}{J. Phys. Chem. B}}
  \textbf{\bibinfo{volume}{113}}, \bibinfo{pages}{6415--6425}
  (\bibinfo{year}{2009}).

\bibitem{Hoover1982}
\bibinfo{author}{Hoover, W.}, \bibinfo{author}{Ladd, A. J.~C.} \&
  \bibinfo{author}{Moran, B.}
\newblock \bibinfo{title}{High-strain-rate plastic flow studied via
  nonequilibrium molecular dynamics}.
\newblock \emph{\bibinfo{journal}{Phys Rev Lett.}}
  \textbf{\bibinfo{volume}{48}}, \bibinfo{pages}{1818--1820}
  (\bibinfo{year}{1982}).

\bibitem{Nose1984}
\bibinfo{author}{Nos{\'e}, S.}
\newblock \bibinfo{title}{A unified formulation of the constant temperature
  molecular dynamics methods}.
\newblock \emph{\bibinfo{journal}{J. Chem. Phys.}}
  \textbf{\bibinfo{volume}{81}}, \bibinfo{pages}{511--519}
  (\bibinfo{year}{1984}).

\bibitem{TuckermanMartyna}
\bibinfo{author}{Tuckerman, A.}, \bibinfo{author}{Lopez-Rendon, J.} \&
  \bibinfo{author}{Martyna, A.}
\newblock \bibinfo{title}{A liouville-operator derived measure-preserving
  integrator for molecular dynamics simulations in the isothermal–isobaric
  ensemble}.
\newblock \emph{\bibinfo{journal}{J Phys A: Math Gen}}
  \textbf{\bibinfo{volume}{39}}, \bibinfo{pages}{5629} (\bibinfo{year}{2006}).

\bibitem{Kuhne}
\bibinfo{author}{K\"uhne, T.~D.} \emph{et~al.}
\newblock \bibinfo{title}{Cp2k: An electronic structure and molecular dynamics
  software package - quickstep: Efficient and accurate electronic structure
  calculations}.
\newblock \emph{\bibinfo{journal}{The Journal of Chemical Physics}}
  \textbf{\bibinfo{volume}{152}}, \bibinfo{pages}{194103}
  (\bibinfo{year}{2020}).

\bibitem{Grimme}
\bibinfo{author}{Grimme, S.}, \bibinfo{author}{Antony, J.},
  \bibinfo{author}{Ehrlich, S.} \& \bibinfo{author}{Krieg, H.}
\newblock \bibinfo{title}{A consistent and accurate ab initio parametrization
  of density functional dispersion correction (dft-d) for the 94 elements
  h-pu}.
\newblock \emph{\bibinfo{journal}{The Journal of Chemical Physics}}
  \textbf{\bibinfo{volume}{132}}, \bibinfo{pages}{154104}
  (\bibinfo{year}{2010}).

\bibitem{ellens2011}
\bibinfo{author}{Ellens, W.}, \bibinfo{author}{Spieksma, F.},
  \bibinfo{author}{Van~Mieghem, P.}, \bibinfo{author}{Jamakovic, A.} \&
  \bibinfo{author}{Kooij, R.}
\newblock \bibinfo{title}{Effective graph resistance}.
\newblock \emph{\bibinfo{journal}{Linear algebra and its applications}}
  \textbf{\bibinfo{volume}{435}}, \bibinfo{pages}{2491--2506}
  (\bibinfo{year}{2011}).

\end{thebibliography}

\end{document}